Part 1:

# Combine Influences of Nanoparticulate Hematite Thin Film Thickness, Roughness, and Weight on Its Photoelectrochemical Performance and Viscous/ Thermal Characteristics of Source Precursor

Romy Loehnert [1, 3], Artur Braun [1], Debajeet K. Bora[1, 2, 4 *1],

[1] Laboratory for High-Performance Ceramics, Empa. Swiss Federal Laboratories of Materials Science and Technology, Uberlandstrasse 129, 8600 Dubendorf, Switzerland

[2] Centre for Nano and Material Sciences, JAIN (Deemed - to - be University), Jain Global Campus, Bengaluru, 562112, Karnataka, India

[3] Department of SciTec, Ernst- Abbe- Hochschule, 07745 Jena, Germany

[4] Laboratory of Inorganic Materials for Sustainable Energy Technologies, University Mohammed VI Polytechnique, Benguerir- 43150, Morocco

[1] Corresponding author: Dr. Debajeet K. Bora, E-Mail: debajeet1@outlook.com; debajeet.bora@um6p.ma

**Abstract**

The objective of this work was to investigate the photoelectrochemical (PEC) performance of nanoparticulate hematite thin film photoelectrodes prepared by a soft- chemistry route. Two cost-effective thin film fabrication techniques were employed to deposit the hematite film. First, the film was deposited on conducting glass substrates by dip coating of the organic precursor containing fatty acid derivatives of iron salts. Process parameters such as the concentration of iron oleic acid derivative precursor solution, the thickness of the organic film, before annealing and the number of deposited layers along with their weight and roughness were studied. In the second approach, the influence of the spin coating process on film formation and respective photoelectrochemical (PEC) performance have been discussed. It was found that the PEC performance of spin-coated samples was lower than that of dip coated samples due to the effect of films rough and smooth characteristics. It is found that the rough surface of the photoelectrode is a prerequisite to achieving good photocurrent values. Here, three-layer samples with roughness between 600nm to 800nm and bulk thickness up to 700nm provided photocurrent densities of 0.6mA/cm$^2$. In part 2 of the manuscript, we have elaborately discussed the roughness and smooth behavior of thin films using X-ray reflectometry technique. Followed by this, a detailed account of the viscous and thermal properties of the fatty acid derivatives of iron precursor have been discussed.

Keywords: Hematite thin film, dip coating, fatty acid derivative of iron salts, Roughness parameters, photocurrent density

# 1. Introduction

In the year 2009, the human population of the world was about 6.8 billion people, and it is expected to climb up to 9 billion people in 2050 [1]. Along with this further industrialization and an increase in living standard and transport will raise the demand for electric power and chemical fuels significantly in the next decades. Today mainly fossil fuels are used for generating the needed energy [2]. But global reserves of fossil fuels are going to be depleted. It leads to the conclusion that other sources of energy must be established. Increasing importance is drawn to clean, renewable energies. Besides wind power and hydroelectricity, the energy of the sun is also an aspect of interest. Since in 1839 Alexandre Edmond Becquerel discovered the photoelectric effect [3] great progress had been made to convert light into usable energy. Today devices like photovoltaic cells and solar thermal systems for private use as well as on an industrial scale are applied to generate electricity and heat. Still, there are more possibilities for solar energy such as the photoelectrochemical (PEC) production of hydrogen. Hydrogen is now considered asone of the essential energy carriers in the future [4]. It has been projected as an essential part of the energy landscape, as the term hydrogen economy was coined in the early 1970s.

The PEC generation of hydrogen is the electrochemical decomposition of water to oxygen and hydrogen at photoelectrodes activated by sunlight. It was first shown in 1972 by Fujishima and Honda using an irradiated semiconducting n-type Titania electrode [8]. In the following years' research intensified and other materials were identified that can be utilized for PEC hydrogen production, for example, tungsten oxide or hematite. Hematite is an attractive material for this application because of its band-gap that enables relatively high absorption of visible light (notwithstanding that hematite has an electronic structure which principally limits its efficiency [9] in comparison to some classical semiconductors); it's thermal and chemical stability and abundance as well as low-cost synthesis [6]. During the 1970s and 1980s, hematite was investigated extensively but found to be a poor photoelectrode due to some of its physical properties. The development and improvement of preparation and characterization techniques allow fabrication and evaluation of materials on the nanoscale now. New approaches to particulate and nanostructured hematite electrodes on this scale are believed to improve the ability of hematite for PEC hydrogen production by facing its physical properties in a better way [7].

A broad variety of different preparation techniques has been used in the recent years to prepare hematite nanoscale films such as spray pyrolysis [8, 9], ultrasonic spray pyrolysis [10], atmospheric pressure chemical vapor deposition [11], sol-gel methods [12], anodic electrodeposition [13] and others [7]. The PEC performance is reported concerning the preparation method and used parameters. In several cases, it could be improved significantly. Factors like morphology and surface area were shown to have a strong influence [10, 13]. Still, there are many preparation and optimization possibilities to provide good hematite PEC photoelectrodes. As a general recipe for materials technology, it is the proper synthesis and processing of a material, which finally brings about the possible theoretical functions of that material in a component or device.

The deposition of films from a liquid solution can be realized with different methods. One of the fundamental, facile and low-cost ways is the dip coating process. In this process, the dipped substrate is vertically removed from the solution entraining liquid in a viscous boundary layer. This boundary layer splits into two parts above the solution surface level. The outer layer returns to the solution; the inner one stays at the substrate forming the film. Right after splitting Landau-Levich-equation can describe the thickness $h_0$ of the film on a substrate from a Newtonian liquid. The thickness $h_0$ of the film on a substrate from a Newtonian liquid.

$$h_0 = 0.94 \frac{(\eta U_0)^{2/3}}{\gamma_{LV}^{1/6}(g\rho)^{1/2}} \qquad (1)$$

Here $\eta$ stands for dynamic viscosity of the solution, $U_0$ is withdrawal speed, $\gamma_{LV}$ is surface tension, g is gravitation constant, and $\rho$ is the density of the solution. Thus, film thickness increases with faster removal and change of solution properties, for example, higher viscosity. With further exposure to the atmosphere, the film is thinned by gravitational drainage and evaporation of solvent leading to a concentration of the nonvolatile constituents and solidification [14].

Another method for film deposition of solutes is the spin coating process. It is as smooth and cost-effective as the dip coating process but provides more uniform films. The spin coating process consists of four steps: First, an excess of liquid solution is deposited on the substrate. Then rotation of the substrate is started. Due to centrifugal forces, the liquid flows radially outward. In the spin-off stage, film thickness becomes uniform over the substrate due to the balance of the two main effects: the rotation-induced centrifugal force that causes radially outward flow and the viscous resisting force acting radially inward [15]. In thicker areas there is a higher mass flow rate, causing dense areas to thin faster till there is uniform film thickness. At the perimeter and substrate edges, liquid accumulates to swellings from where drops flung off. Evaporation that takes place during the whole process then takes over as the fourth and main thinning step. The centrifugal outflow of the solution is stopped due to raised viscosity of the concentrated solution. An equation found by Meyerhofer can estimate final film thickness $h_f$ that also takes evaporation of solvent into account.

$$h_f = \frac{C_0}{\Omega^{1/2}} \left( \frac{3k\eta_0}{2(1-C_0)\rho} \right)^{1/3} \qquad (2)$$

Here $C_0$ is the initial concentration of the solution, $\Omega$ is the angular rotation rate, $\eta_0$ is the initial dynamic viscosity, $\rho$ is the density of the fluid and k is a solvent dependent constant for evaporation. Thus, thicker films are obtained with increasing viscosity of the solution, solids concentration, volatility of solvent or by decreasing the spinning speed $\Omega$ [16, 17].

As evaporation rate in both processes is fast, the solute has little time to order. With higher viscosity of solvent evaporation and more rapid concentration can influence microstructure formation [14].

The objective of the current study is to investigate the PEC performance of hematite nanoparticulate films processed by a soft chemistry preparation route. The preparation route has already been described elsewhere to fabricate hematite photoelectrodes using the dip coating method. A maximum photocurrent of 0.45mA/cm$^2$ could be measured for a four-layer sample with a thickness of about 600nm [18]. The photocurrent density obtained in this case is influenced by several factors of film microstructure formation such as precursor solution viscosity, use of different fatty acids, thin film weight and roughness. The central theme of the current study is the investigation of the film formation details obtained by dip and spin coating withstanding this fact.

## 2. Materials and Methods

All chemicals Iron Nitrate nonahydrate, oleic acid, stearic acid and lauric acid, THF used for the synthesis of hematite films are of analytical grade with purity of 99.99 % and acquired from Sigma-Aldrich, Switzerland.

### 2.1 Synthesis of precursor with fatty acids having different chain lengths

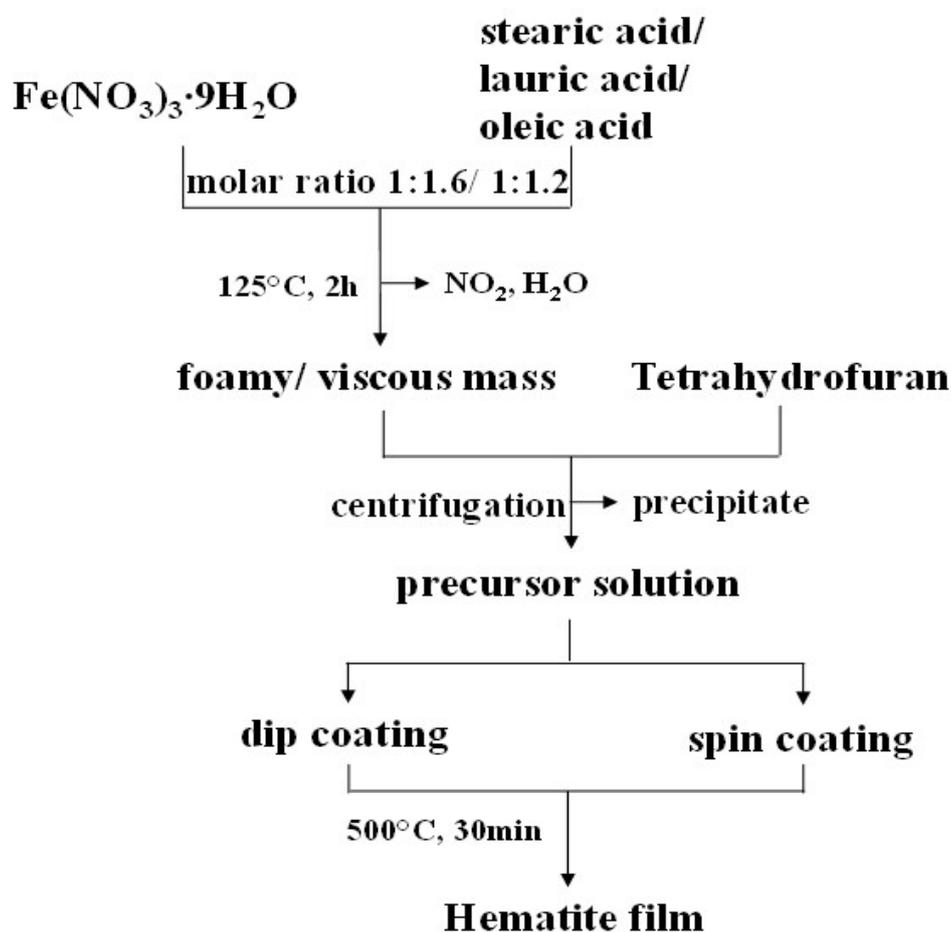

**Scheme 1:** Synthesis route for making hematite film utilizing both dip coating and spin coating process

A non-aqueous precursor based on oleic acid was prepared (see scheme 1) for synthesizing hematite nanoparticulate films. 17g of oleic acid ($C_{18}H_{34}O_2$) were heated up in a beaker to 125°C on a hot plate. Gradually 28g of iron (III) nitrate (Fe $(NO_3)_3 \cdot 9H_2O$) were added under constant mechanical stirring leaving coloring of the solution turns dark reddish brown immediately. After five to ten minutes of heating, brownish $NO_2$ gases and water vapor started to evolve turning the solution into a foamy mass. It was mechanically stirred for two hours at a temperature of 125°C.

With the evolution of $NO_2$ gases, the reactant mass got more viscous. When no additional gases formed, it was cooled down at room temperature for 24 hours to gain a shiny, dry mass (see figure 1) that was reduced to smaller pieces.

By adding tetrahydrofuran ($C_4H_8O$, THF) and stirring for 20 to 30 minutes a solution was formed, that was sonicated to homogenize and decanted to separate from undissolved parts. Further precipitate could be separated by five minutes lasting centrifugation at acceleration equal to 2600 times gravity and decantation of the solution.

The fraction of iron oleate complex in the precursor solution was determined by weight per volume concentration (w/V concentration). It is defined as

$$\frac{w}{V}[\%] = \frac{m(solute)[g]}{V(solution)[ml]} \cdot 100 = \frac{m(solution) - m(solvent)[g]}{V(solution)[ml]} \cdot 100 \qquad ()$$

The obtained w/V concentrations are in the range of 3.4% to 4.8% (see table 1). The solution was used as a stock solution. Changes in w/V concentration were carried out by adding THF to the stock solution to investigate the influence of different concentrations of the precursor on the PEC performance,

*Table 1: Weight per volume concentrations for three precursors*

| Precursor | Precursor 2 | Precursor 3 | Precursor 4 |
|---|---|---|---|
| w/V concentration | 3.4% | 4.2% | 4.8% |

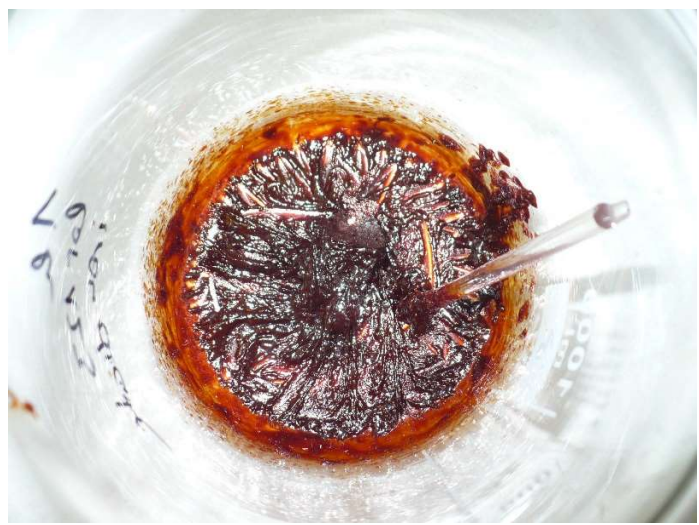

**Figure 1**.: Heat treated and cooled down the shiny, dry mass of reaction product of iron nitrate and oleic acid.

## 2.2  Preparation of precursor with stearic acid and with lauric acid

Synthesis of a non-aqueous precursor based on stearic acid was done according to the method of Deb and Basumallick [19]. Stearic acid ($C_{18}H_{36}O_2$) and iron (III) nitrate nonahydrate ($Fe(NO_3)_3 \cdot 9H_2O$) in a molar ratio of 1:1.6 were used. The first 17g of stearic acid were molten in a beaker on a hot plate that showed 125°C. 28g of iron nitrate were added gradually under constant mechanical stirring, coloring the solution reddish brown following the same procedure as used in the case of oleic acid.

Similarly, Lauric acid precursor is based on the same preparation procedure. 12g of lauric acid ($C_{12}H_{24}O_2$) was molten at a temperature of 125°C. 28g of iron nitrate was gradually added under mechanical stirring, equal to a molar ratio lauric acid to hydrated iron nitrate of 1:1.6. After 20 minutes of tiny heating bubbles started to evolve and 30min, later the solution turned into a foamy mass. The foam was stirred mechanically for one more hour at a temperature of 125°C. When the gas evolution decreased, it was cooled down at room temperature for 24 hours to form a soft waxy reddish mass. It was reduced to smaller pieces and treated like the oleic acid precursor.

The obtained weight per volume concentrations of the supernatants is listed in table 2.

Concerning the preparation of iron oleate precursor, the reaction and formation of bubbles during heating are more sluggish. The formation of a highly viscous mass during development could not be observed. The achieved w/V concentrations of precursors are low. For a slightly longer heat-treated precursor a precipitate formed even after centrifugation.

Table 2: Weight per volume concentrations of iron stearate and iron laurate precursors

| Precursor solution | Precursor 5 (stearate) | Precursor 6 (stearate) | Precursor 7 (laurate) |
|---|---|---|---|
| w/V concentration | 1.2% | 2.0% | 2.8% |

## 2.3 Hematite thin films prepared by dip coating.

Films of hematite were deposited on soda lime glass substrates with a conductive layer of fluorine-doped tin oxide (FTO, 30mm x 12mm x 2.3mm, *Pilkington TEC 8*) by dip coating. Studies concerning the concentration of the precursor solution, concerning the weight of the deposited organic film and the number of deposited layers were performed.

For concentration-dependent studies, diluted solutions of iron oleate precursor 2 and precursor 3 were used. The FTO substrate was immersed ten seconds in the precursor solution and removed slowly within eight seconds. The sample was then dried for ten minutes on a hot plate at a temperature of 72°C. The coated samples were kept in the air-vented\ furnace in a non- isothermal manner at 500°C. They were annealed for 30 minutes at that temperature in air and then taken out of the furnace and cooled down at room temperature. The annealing temperature and time have already been optimized in earlier works using iron oleate precursor solution for hematite photoelectrode fabrication [18]. Several layers were deposited on the substrate (see table 3, table 4). For this the cooled down sample was again coated, dried and annealed for half an hour at 500°C.

Table 3: Number of layers for dip coated samples prepared with different concentrated solutions of precursor 2

| precursor 2 w/V concentration | 0.15% | 1% | 2% | 3% | 3.4% |
|---|---|---|---|---|---|
| number of layers | 6 | 1 | 4 | 4 | 4 |
|  | 12 | 4 | 8 | 8 | 8 |

|  | 15 | 8 |  |  |  |

Table 4: Number of layers for dip coated samples prepared with different concentrated solutions of precursor 3

| precursor 3 w/V concentration | 1% | 1.5% | 2% | 2.5% | 3% | 3.5% | 4.2% |
|---|---|---|---|---|---|---|---|
| number of layers | 4 | 4 | 4 | 4 | 4 | 4 | 4 |

For weight dependent studies, the cleaned FTO substrate was weighed and then covered with tape to leave an uncovered area of $12 \times 11 mm^2$ that is coated. The substrate was immersed for two minutes in the precursor solution, removed in about one second and then dried for ten minutes on a hot plate at a temperature of 72°C. Afterward, the tape covering was removed, and the weight of the coated substrate was measured. By subtracting the initial weight of the substrate, the film weight was determined. The weight of organic film was measured as a proportional value to its thickness.

To raise the thickness of the organic film the substrate was covered again and further material was deposited by immersion for only one to two seconds, fast withdrawal speed and drying for ten minutes on a hot plate at 72°C. If the substrate was immersed longer into the precursor solution the already deposited film got dissolved, and there was no increase of organic film weight. For weight dependent studies the as received concentrations of iron oleate precursors as well as of iron stearate and iron laurate precursor were used and weights in the range of 1.5mg to 12mg were deposited. The average weight of organic film that is obtained with several numbers of dips dependent on the w/V concentration of precursor solution is shown in table 5. The weight of the sample can be reproduced quite well by keeping the deposition parameters.

Table 5: Averaged weight of organic film dependent on the number of dips and weight per Volume concentration of used precursor
(* concentration could be higher due to evaporation of solvent)

|  | 1 dip | 2 dips | 3 dips | 4 dips |
|---|---|---|---|---|
| 1% | 1.3mg | 1.8mg | 2.8mg | 3.5mg |
| 2% | 1.3mg | 2.0mg | 3.0mg | 4.5mg |
| 3% | 1.3mg | 3.1mg | 4.9mg | 6.8mg |
| 3.5% | 1.5mg | 3.0mg | 5.0mg | 7.4mg |
| 4.2% (*) | 2.3mg | 5.7mg | 10.0mg | No perform |
| 4.8% | 1.7mg | 3.7mg | 6.7mg | 10.6mg |

The dried organic film of iron oleate precursor is reddish brown, sticky and highly viscous like described elsewhere [20]. Iron stearate and iron laurate films solidified after drying to a waxy mass. The coated samples were annealed in the same way as samples of concentration studies. If the sample showed a film with peeling off the surface after annealing, it was softly wiped off to remove loose parts. In the studies concerning some deposited layers, further layers were deposited on different samples of weight dependent studies. Samples of precursor 3 were covered with an organic film of the same weight (3.8mg) and annealed to get two-layer samples. On samples of precursor 4, a second and also a third layer was deposited. Here the weight of deposited layer also the same for all samples (3.6mg).

In general, the organic film thickness of dip coated samples is not equal over the substrate due to the deposition method. When the substrate is withdrawn, it contains liquid. Besides evaporation of solvent also some of the liquid drains off the substrate and accumulates at the bottom, leading to the greater thickness of organic coating in this area (see figure 2).

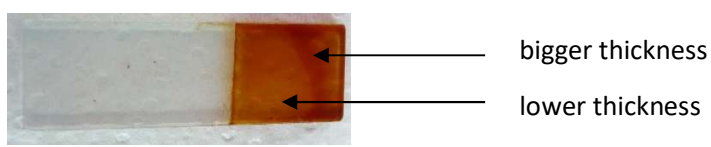

**Figure 2**: Thickness difference between bottom and center of the film due to the draining of solution after withdrawal

## 2.4 Hematite films prepared by spin coating

The iron oleate precursor solution was deposited by spin coating on FTO substrate. The cleaned substrate was covered with tape to fix an area of 12x11mm² that is coated. The substrate was placed on the sample holder of the spin coater (*Primus STT25*) with the area to be coated centered and fixed to it by under pressure. The precursor solution was dispensed over the open area with a pipette (maximum volume: 3ml). The amount of used precursor was six drops. The substrate was spun up at 200rpm for two seconds and then spun at higher angular speed for one minute. Different angular speeds were used to get different film thicknesses of an organic film (see table 6). The coating was then dried for 10 minutes at 72°C on a hot plate, and the weight of organic film was measured.

Spin-coated samples show uniform film thickness in the middle of the substrate but the accumulation of precursor on the edges, the amount dependent on the angular rate and viscosity of precursor solution (see figure 3). Reproduction of organic film weight using the same deposition parameters is difficult. Usage of precursor solution is higher compared to dip coating. Spin-coated samples were annealed at 500°C in the air similar to dip coated ones.

*Table 6: Weight of organic film dependent on w/V concentration of used precursor and angular speed*

*(\* concentration could be higher due to evaporation of solvent)*

| precursor solution, w/V concentration | angular speed [rpm] | the weight of the organic film |
|---|---|---|
| prec3, 3% | 2000 | 1.8mg |
| prec3, 3% | 1000 | 3.8mg |
| prec3, 3% | 750 | 6.9mg |
| prec3, 3.5% | 750 | 4.51mg |
| prec3, 4.2% (*) | 750 | 8.46mg |
| prec3, 4.2% (*) | 600 | 13.37mg |
| prec3, 4.2% (*) | 500 | 29.70mg |
| prec4, 4.8% | 700 | 4.52mg |
| prec4, 4.8% | 600 | 6.37mg |
| prec4, 4.8% | 500 | 8.36mg |

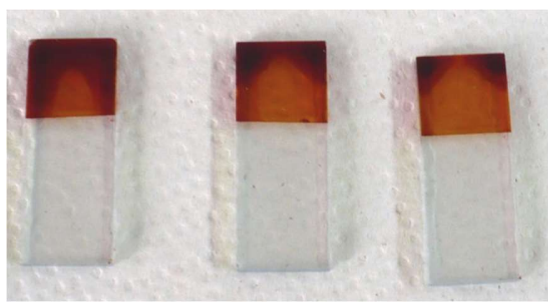

**Figure 3**: Thicker accumulation (inhomogeneity) of precursor solution at the edges of the substrate with higher spin speed, spin speed is 500rpm, 600rpm and 700rpm from left to right, respectively.

## 2.5 Metrology and diagnostics

The rheological parameters of some precursor solutions were investigated with a rotational viscometer Viscotherm VT 100 by Physica using a double gap measuring system according to DIN 54453. The gap size is 0.5mm. The torque, shear stress τ, and dynamic viscosity η were measured as a function of applied shear rate $\dot{\gamma}$ in the range of $1s^{-1}$ to $1300s^{-1}$ at a temperature of 23°C. As the viscosity of precursor is low, the dual gap system is needed to ensure laminar flow conditions for reliable measurement results.

Also, some parts of precursor solution of iron oleate precursor were dried on a hot plate that showed a temperature of 70°C till a mass similar to the dried film deposited on FTO substrates had formed. The dried precursor was investigated with simultaneous thermal analysis (STA) using a Mettler Toledo TGA/SDTA851 machine. The sample was heated up in an alumina vessel from room temperature to 600°C at a heating rate of 1K/min. Weight change (TG) and temperature difference to an alumina reference sample (DTA) were measured and normalized to the weight of the sample.

### 2.5.1 Determination of photoelectrochemical performance

Linear sweep voltammetry under dark and light condition was performed to determine the PEC performance of hematite films. The sample used here is the working electrode in a three-electrode configuration. A platinum wire is set as a counter electrode, and an Ag/AgCl/3 M KCl electrode is used as a reference electrode. The electrodes are immersed in 1M KOH electrolyte (see figure 4a). Electric contact to the hematite film is done via the FTO coating on the glass substrate and a titanium crocodile clip. In light condition, an area of $0.6cm^2$ of the sample is illuminated with simulated sunlight of AM 1.5 global standard solar spectrum [21] generated by a 1 Sun Oriel Lamp by L.O.T – Oriel AG. Wavelengths lower than 280nm and higher than 800nm are removed from the spectra with a fused quartz glass filter. For the measurement, a potentiostat (Voltalab80 PGZ 402 by Radiometer analytical) is used. The setup is shown in figure 4b.

In dark condition, there is no electrochemical reaction at the hematite anode until a voltage is reached that enforces the electrolysis of water. The current density increases steeply above this potential in the voltammetry pattern. In light condition, electron-hole pairs are generated in the hematite photoanode. If the generated photovoltage is high enough and the applied bias shifts the counter electrode potential above the potential of H/H$^+$ redox pair, decomposition of water is induced already at applied voltages lower than that for the dark condition.

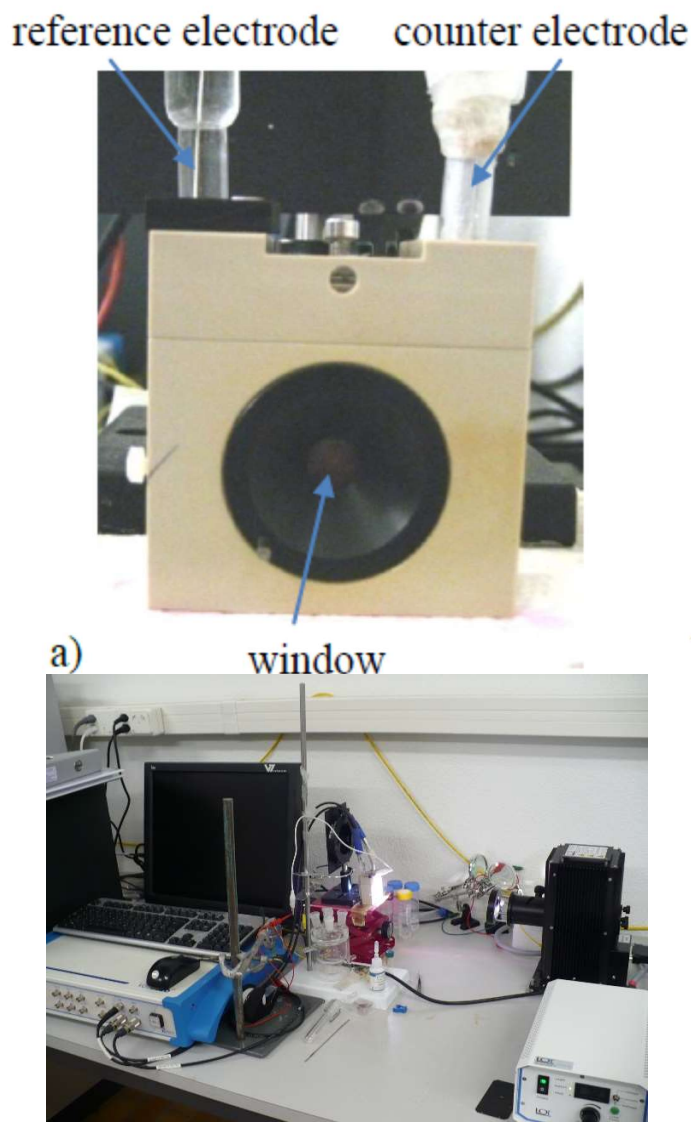

**Figure 4**: (a) A cappuccino cell with hree-electrode configuration with illuminated hematite film as the working electrode and (b) Complete measurement setup.

For the measurement, the open circuit potential of the sample was measured first for four minutes to determine the rest potential. Then the current density J is measured as a function of the applied voltage V. The measurement was started at open circuit potential of the sample and applied potential was changed up to 600mV at a speed of 10mV/s. The measurement was done in dark condition first and then repeated with the illumination of the sample. The error of measured current density is $0.2\mu A/cm^2$ below a range of $120\mu A/cm^2$, and above this, it is $2\mu A/cm^2$. The value of achieved photocurrent density is taken at an applied voltage of 0.42V, equal to the voltage of water electrolysis against RHE for the used measurement conditions.

### 2.5.2 Profilometry for film thickness and roughness determination

Film thickness and roughness of the prepared samples were measured with a mechanical profilometer (XP-1 Stylus Profiler by Ambios Technology). A stylus with a conical diamante tip with a diameter of 5µm is in contact with the sample surface and is moved laterally over it. Changes in thickness and roughness of the sample induce a vertical displacement of the stylus that is recorded and converted to a digital signal. From these data, the unfiltered profile is displayed [22]. Before measurement, the film was scratched twice with a knife at an interval of 5mm in the area that is illuminated at the analysis of PEC performance (see figure 5). The profile between these two scratches was measured with a scan speed of 0.05mm/s and a stylus force equal to 0.5mg. The scratches were used as reference points to define the level of the substrate in the Ambios XP-2 evaluation software. An error in leveling may occur due to the roughness of substrate ($R_a \approx 25nm$) and plastic deformation of FTO layer during scratching. The minimum horizontal resolution is 15Å, and the vertical resolution is 200nm for the used settings.

Roughness parameters root mean square $R_{rms}$ and arithmetic average $R_a$ were calculated by the evaluation software from the raw profile. Also, the arithmetic average $\bar{h}$ and median $\tilde{h}$ of thickness as well as the height histogram were calculated from the data. For the height histograms, the thickness data points were binned in groups with a binning width of 50nm. The first class has a value of -50nm and contains all height data x with $-75nm \leq x < -25nm$. The waviness of the profile was not corrected and may influence the calculated values.

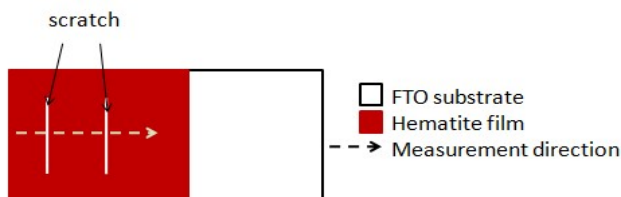

**Figure 5**: Sketch of sample preparation for profile measurement

## 3. Results and discussion

### 3.1  Viscous and thermal properties of the fatty acid derivative of an iron salt precursor

The produced precursor solutions based on oleic acid showing different weight per volume concentrations w/V were investigated to study their viscosity and thermal properties. The precursor solutions show a nearly constant dynamic viscosity η with increasing shear rate $\dot{\gamma}$ and thus seem to behave like a Newtonian liquid. The dynamic viscosity of precursor solution increases with a higher weight per volume concentration (see figure 6).

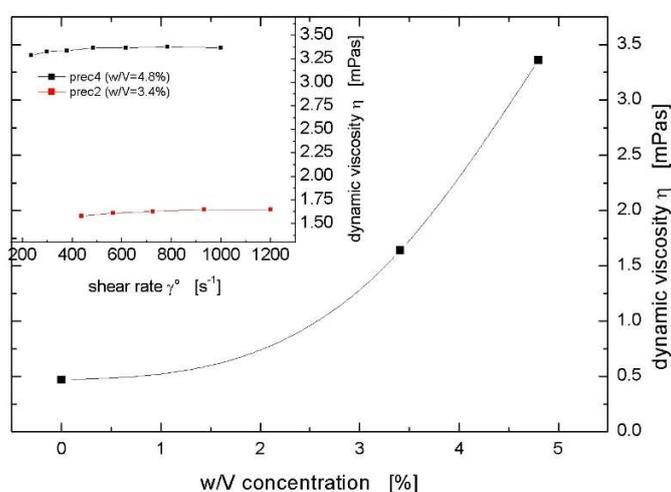

**Figure 6**: Dynamic viscosity η dependent on w/Concentration of the iron oleate precursor solution and a constant dynamic viscosity at different shear rates $\dot{\gamma}$ of different precursor solutions (inset).

Thermal decomposition of dried iron oleate precursor and of the precipitate that was collected at centrifugation was investigated with simultaneous thermal analysis. The precipitate is found to decompose likewise the iron stearate complex investigated by Deb and Basumallick [19] with starting temperatures and peaks shifted about 10°C to lower temperatures (see figure 7a). The main decomposition takes place in a temperature range between 200°C and 350°C. The small DTA peak with an onset temperature of 419°C that is not combined with a change of mass indicates the transformation of γ-$Fe_2O_3$ nanoparticles to α-$Fe_2O_3$ structure [18].

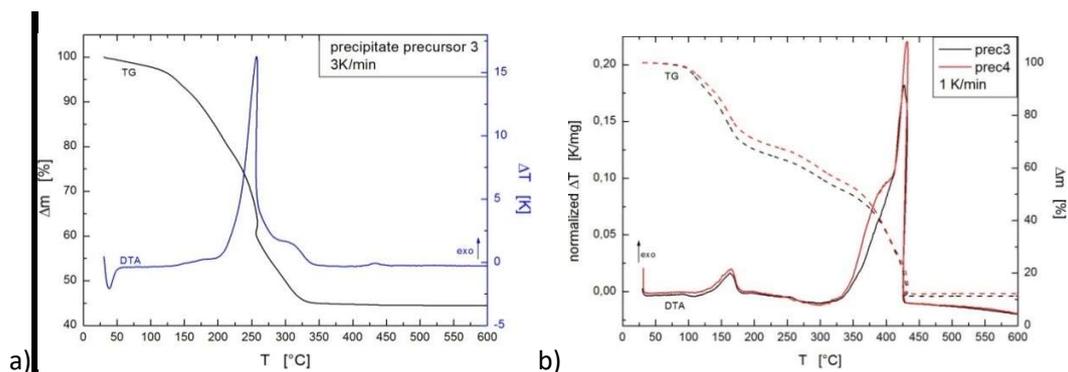

**Figure7**: Thermal decomposition behavior of (a) iron oleate precipitate and (b) dried iron oleate precursor

In contrast, dried parts of iron oleate precursor 3 and 4 show a first exothermic peak combined with a bigger weight loss between 125°C to 180°C. This peak could be due to residual iron nitrate that has not decomposed or reacted during preparation and is dissolved in the precursor solution. Another hint for further reaction of iron nitrate is the formation of tiny bubbles during drying of the dip-coated layer that could sometimes be seen. Also, pure iron nitrate shows several steps of thermal decomposition, the main around 150°C dependent on the heating rate [23].

The primary thermal precursor decomposition starts at a temperature of around 340°C when the dissolution of precipitate is about to finish. It also shows a two-step process. But concerning precipitate, the first peak has a lower temperature difference signal than the second step (see figure 7b). The total weight loss of dried precursor is 1.6 times higher than that of the precipitate. The reason for this could be that besides iron oleate complex also residual iron nitrate and oleic acid are dissolved in the solvent and therefore still in the precursor solution. The excess residual oleic acid in the precursor could be the reason for the shift of the primary thermal decomposition to higher temperatures. Precipitate prepared with the higher amount of stearic acid shows similar decomposition behavior [19]. Iron oleate complex is dissolved in the excess oleic acid and decomposition behavior in solution could be different compared to the dry state of the precipitate. Also, a difference in the decomposition of different precursors can be seen. The preparation of precursors only differs in time of thermal treatment at 125°C as the heating was stopped after no more bubbles evolved and a viscous mass had formed. Precursor 3 shows a weak endothermic peak starting at a temperature of 100°C that cannot be found with precursor 4. The difference is even more significant to see in the TG curve as precursor 3 loses more weight from this onward temperature. The first exothermic peak seems to be similar for both precursors and also the endothermic DTA signal starting at 250°C is the same. But there is a difference at the primary decomposition. Precursor 4 shows a better-developed peak shape than precursor 3 at the first step and also more significant weight loss. Furthermore, a higher temperature difference at the peak of the second step is measured. There could be a higher fraction of organic species in dried precursor 4 that reacts and develops more heat along. Oleic acid has an evaporation temperature of 360° C. This temperature fits well with the position of the first step of the primary decomposition peak. Therefore, the first peak could be due to the decomposition of oleic acid. As there is no further change after the main exothermal peak up to the temperature of 600°C, the iron oleate structure should decompose, and hematite phase should form in this step.

### 3.2 PEC performance and thickness properties of dip-coated hematite thin films from an oleic acid derivative of an iron precursor

Concentration-dependent studies

Studies with different concentrations of precursor solutions and with varying thicknesses of the deposited organic film as well as studies concerning the number of deposited layers were done to investigate PEC performance. Annealed films of concentration study are orange translucent, shiny and smooth (see figure 8a). They show a roughness of 25nm to 70nm. The thickness of one-layer samples is in the range of 3nm to 50nm. With the use of higher concentrated precursor solution, the obtained thickness is more significant. This increase was found to be by Landau-Levich equation for dip coating method: With higher w/V concentration also viscosity of precursor is higher. Increased viscosity of the solution leads to more top film thickness at the dip coating process [24].

With the deposition of further layers on the annealed film, the total film thickness grows (figure 9). The surface is still flat; roughness does not change. Along with this, the color of the film turns to reddish-brown. Flat films of higher thickness show no difference between the current density of dark and light. Only a low photocurrent density in the range up to 30μA/cm2 can be observed on films of thickness around 10nm to 40nm.

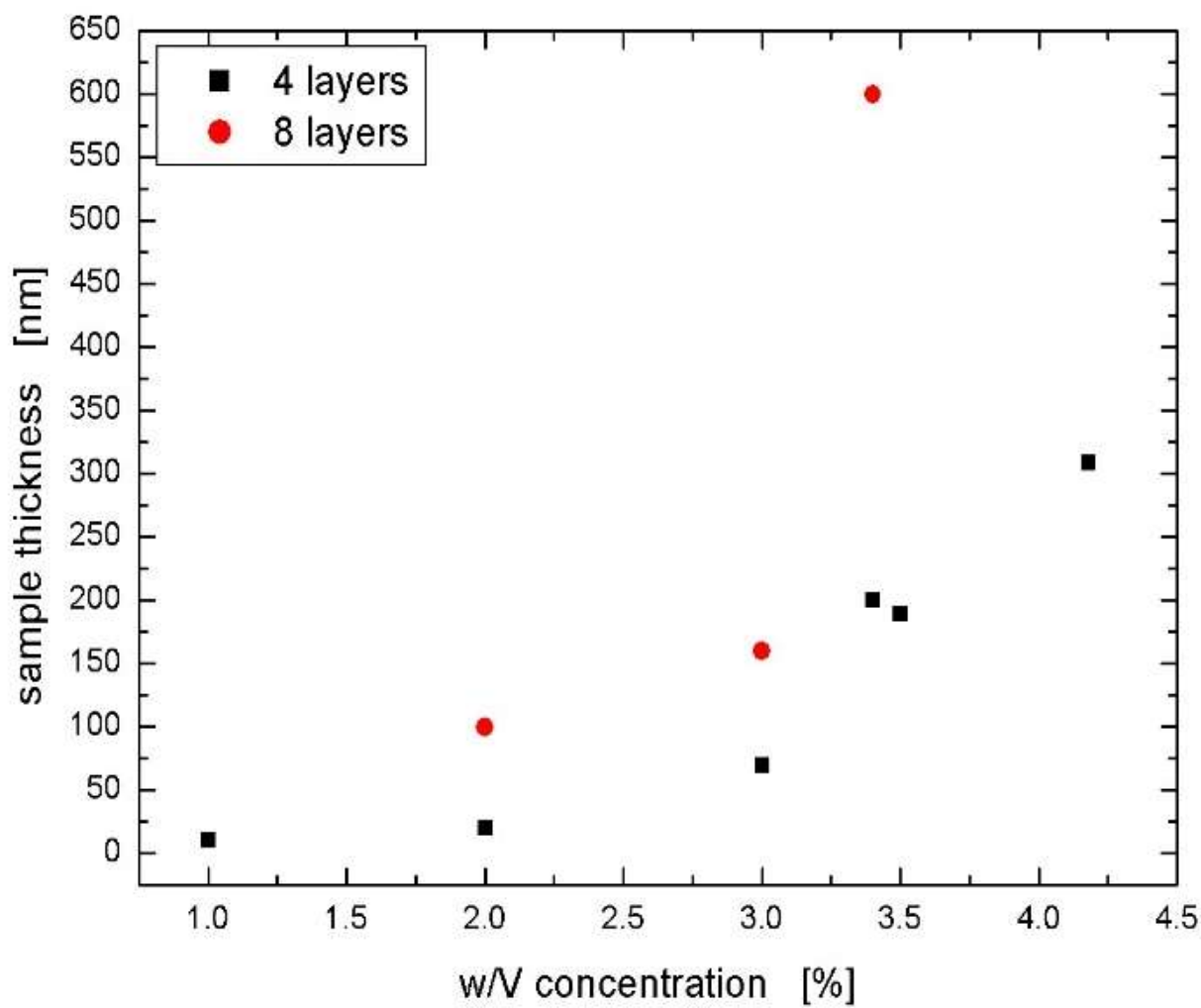

**Figure 8**: Film thickness of smooth samples dependent on w/V concentration of precursor solution and number of deposited layers (black squares 4 layers; red circles 8 layers).

*Weight dependent studies*

For weight dependent studies substrate removal speed was fastened, and higher concentrated precursor solutions were used. Along with Landau- Levich equation thickness of these deposited organic films is higher than for organic films of concentration studies. Instead of thickness, the weight was measured as a proportional value. Annealed samples with a higher weight of organic film show cracks besides also the smooth surface (see figure 9b) and peeling off the material leaves a rough surface after wiping (see figure 9c). There is no smooth surface with a further increase in organic film weight above 1.5mg.

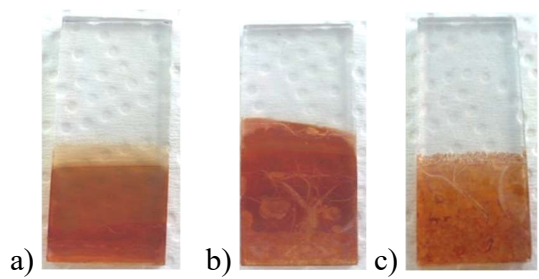

**Figure 9**: Surface of annealed samples with (a) smooth, (b) smooth with cracks and (c) rough surface.

Most samples of weight dependent studies have a rough surface. One-layer samples with a rough surface show an enhanced PEC performance that correlates with the weight of the organic film. Up to 6mg of organic film weight, photocurrent density increases. A maximum is found at 6- 7 mg. Photocurrent density decreases (see as example prec 3 figure 10) at a higher weight. In the region below 5 mg, photocurrent density values show high scattering (see figure 10 – prec 3 and prec 4). Also, there is a difference in the PEC performance of one-layer samples with rough surface dependent on the used precursor (see figure 10). Samples of precursor 3 show the best performance. The value of maximum photocurrent density is 528µA/cm$^2$. Samples of all other used precursors display photocurrent densities in lower ranges. Especially precursor solutions 3 and 4 have been investigated.

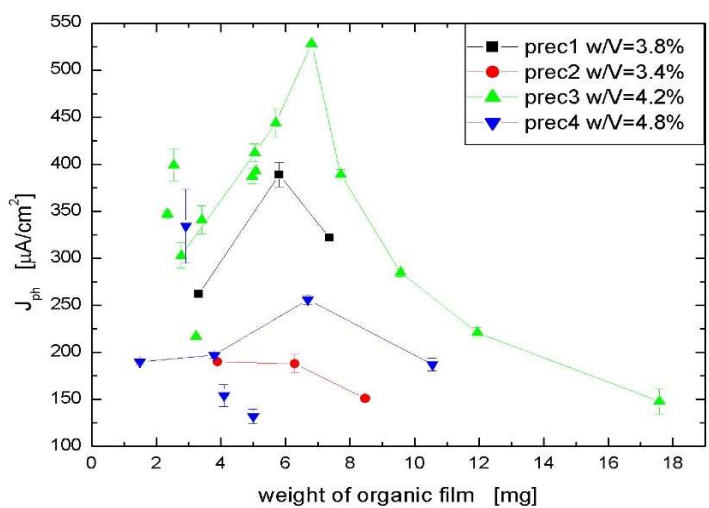

**Figure 1**: Photocurrent density of samples dependent on the weight of the initial organic film

A relation between the weight of organic film and the thickness or roughness of the annealed sample does not exist. The ranges of thickness and roughness values for rough one-layer dip coated samples of weight dependent studies are shown in table 7. By comparison of these values, one can find that the median is nearly the same for samples of the different precursors. The median can be said to represent bulk thickness as this statistic average value is not influenced by outliers like the roughness peaks are. The median of the samples is mainly restricted to a range from 100nm to 250nm (see figure 11a). Samples of precursor 1 show a slightly less and samples of precursor 2 a marginal bigger median than samples of precursor 3 and 4. The photocurrent density for all samples is scattered all over this range of median (see figure 11a). Each precursor solution by itself leads to a certain range of roughness $R_{rms}$. Most of the samples are within a range of $R_{rms}$ of 500 nm to 800 nm. Like for the median, the values of photocurrent density scatter over this restricted range (see figure 11b). Precursor 1 and precursor 3 that show the highest amounts of photocurrent density display the same range of roughness, 600nm to 800nm (see table 7). The roughness of the sample influences the value of the arithmetic average of thickness. The ranges of the arithmetic average coincide for precursors 1, 3 and 4. The range is about 350 nm to 500 nm. For samples of precursor 2, averages and roughness values are higher. Also, here the PEC activity of samples scatters all over the range (see figure 11c). Neither for the single thickness and roughness values nor combinations of them a dependency of photocurrent density, like for weight of the organic film, can be found (see figure 11). The deposition of the same weight of organic film using a diluted precursor solution leads to thinner and less rough samples after annealing. Also, the PEC performance is less.

*Table 7: Ranges for calculated thickness and roughness values of rough one-layer samples of different precursors*

| precursor | the range of $\tilde{h}$ [nm] | the range of $\bar{h}$ [nm] | the range of $R_{rms}$ [nm] |
|---|---|---|---|
| 1 | 80 - 150 | 330 - 550 | 600 - 800 |
| 2 | 170 - 360 | 500 - 800 | 700 - 1500 |
| 3 | 100 - 250 | 250 - 600 | 600 - 800 |
| 4 | 160 - 250 | 350 - 500 | 350 - 550 |

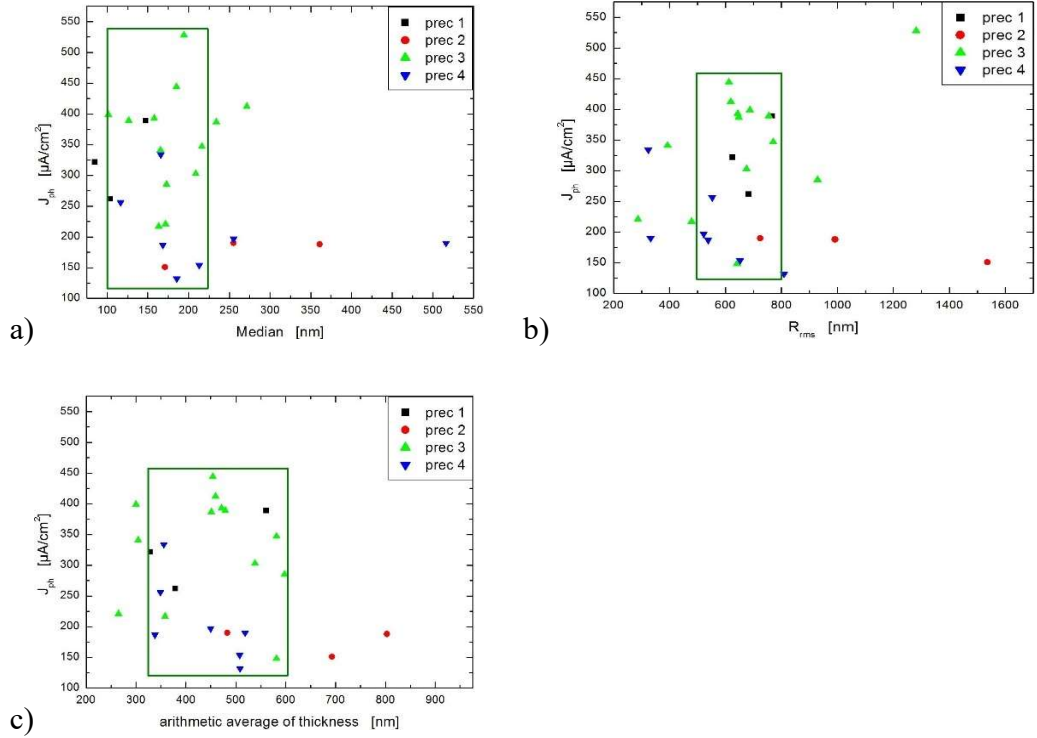

**Figure 11**: Photocurrent density of rough one-layer samples as a function of median (a), of roughness $R_{rms}$ (b) and of arithmetic average of thickness (c), the rectangles indicate the range of median, $R_{rms}$ and average, respectively, in which most samples can be found and also that there is no dependency of photocurrent density on median, roughness, arithmetic average, respectively.

For studies concerning the number of deposited layers, the samples of weight dependent study of precursor 3 and precursor 4 were used. With the deposition of a second layer, the photocurrent density increases further. The samples are less translucent and more reddish brown. For precursor solution 4, samples display now a photocurrent density in the range of 300μA / cm$^2$ to 400 μA / cm$^2$ (see figure 12b). For samples of precursor 3, the photocurrent density increases up to around 600μA / cm$^2$ (see figure 12a). A small maximum can still be found for samples that also displayed a maximum in photocurrent density with one layer of weight dependent studies. For samples of low weight of organic film for the first layer, the achieved new PEC performance is still less than for other two-layer samples. It is found that photocurrent density for two-layer samples is narrower than for one-layer samples.

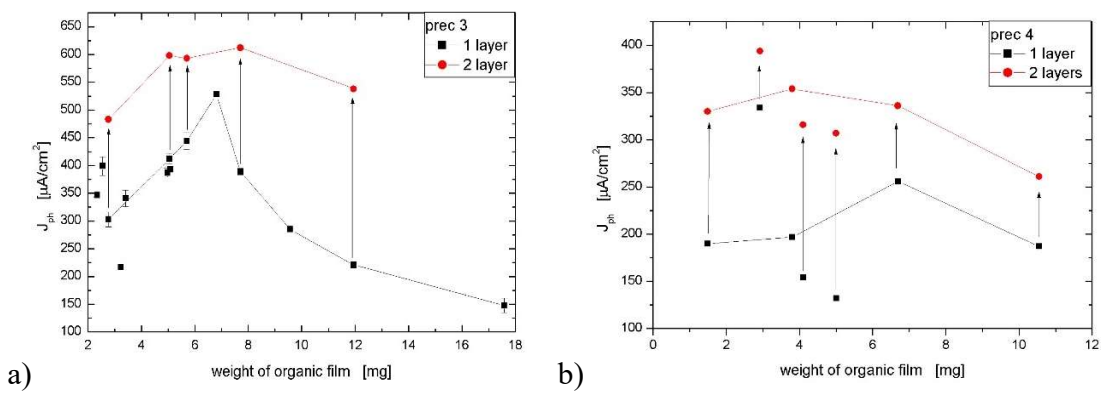

**Figure 12**: Photocurrent density for one- and two-layer samples of precursor 3 (a) and precursor 4 (b), the arrows indicate the increase of photocurrent density for each sample

The thickness and roughness of two-layer samples increase compared to one-layer samples. The maximum roughness peaks are still at the same height but the number of smaller peaks increases. It can be seen by comparison of the profiles of a sample with one and with two layers (see insets figure 13 a, b). Also, the thickness histograms reveal this change. The frequency of thicknesses higher than the bulk level of 500nm is bigger than for the one-layer sample (see figure 13).

The thickness histograms also prove an increase of bulk thickness. For one-layer samples, there is a main narrow peak with high frequency at thicknesses of 100 nm to 200 nm. It is changed to a broad peak with lower maximum frequency at 200 nm to 300 nm for two-layer samples (see figure 13). Therefore, the fraction of higher bulk thicknesses is bigger in the two-layer sample. Also, the median is shifted for the two-layer samples to a range of 330 nm to 500 nm (see table 8).

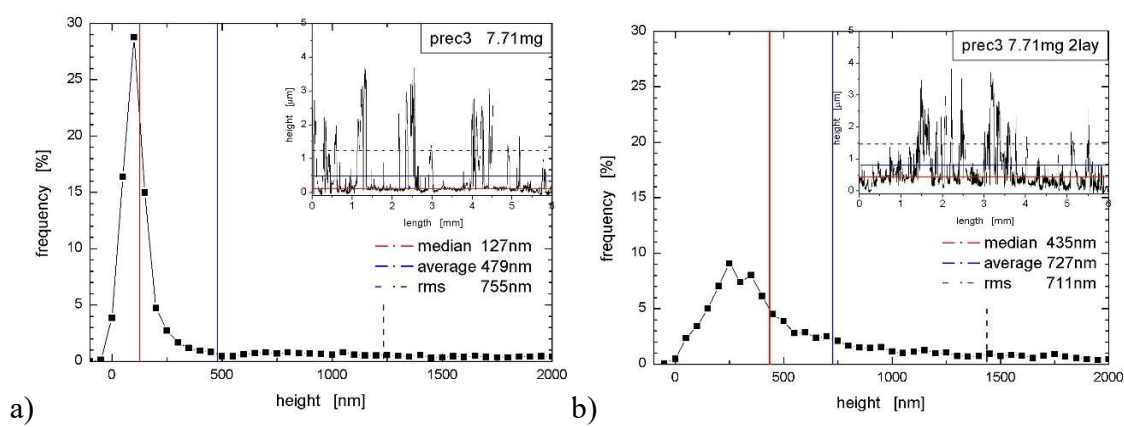

**Figure 13**: Thickness histograms and profiles for one layer (a) and two layers (b) of sample preparation of precursor 3.

For samples of precursor 3, the values of R$_{rms}$ only change slightly and still are in the same range as one-layer samples (see table 8), only with little variation. As the thickness histograms show a change in roughness, the R$_{rms}$ value seems not to be significantly affected by a higher number of smaller peaks. On the contrary for two-layer samples of precursor 4, the roughness is shifted from lower values for one-layer samples to the same range as samples of precursor 3 shows. The average is also changed to higher values and ranges of precursor 3 and 4 still coincide.

*Table 8: Ranges for calculated thickness and roughness values of rough samples with a different number of layers*

|  | range of $\tilde{h}$ [nm] | | range of $\bar{h}$ [nm] | | range of R$_{rms}$ [nm] | |
| --- | --- | --- | --- | --- | --- | --- |
|  | Prec 3 | Prec 4 | Prec 3 | Prec 4 | Prec 3 | Prec 4 |
| 1 layer | 100 - 250 | 160 - 250 | 250 - 600 | 350 - 500 | 600 - 800 | 350 - 550 |
| 2 layers | 330 - 400 | 330 - 500 | 600 - 780 | 550 - 900 | 600 - 800 | 600 - 700 |
| 3 layers |  | 500 - 700 |  | 800 - 1000 |  | 600 - 1000 |

With the deposition of a third layer on samples of precursor 4 of weight dependent studies, the PEC performance still increases. Only one sample shows a decrease (see figure 14). The range of achieved photocurrent densities is 450μA/cm$^2$ to 600μA/cm$^2$ and coincides with the range of two-layer samples of precursor 3.

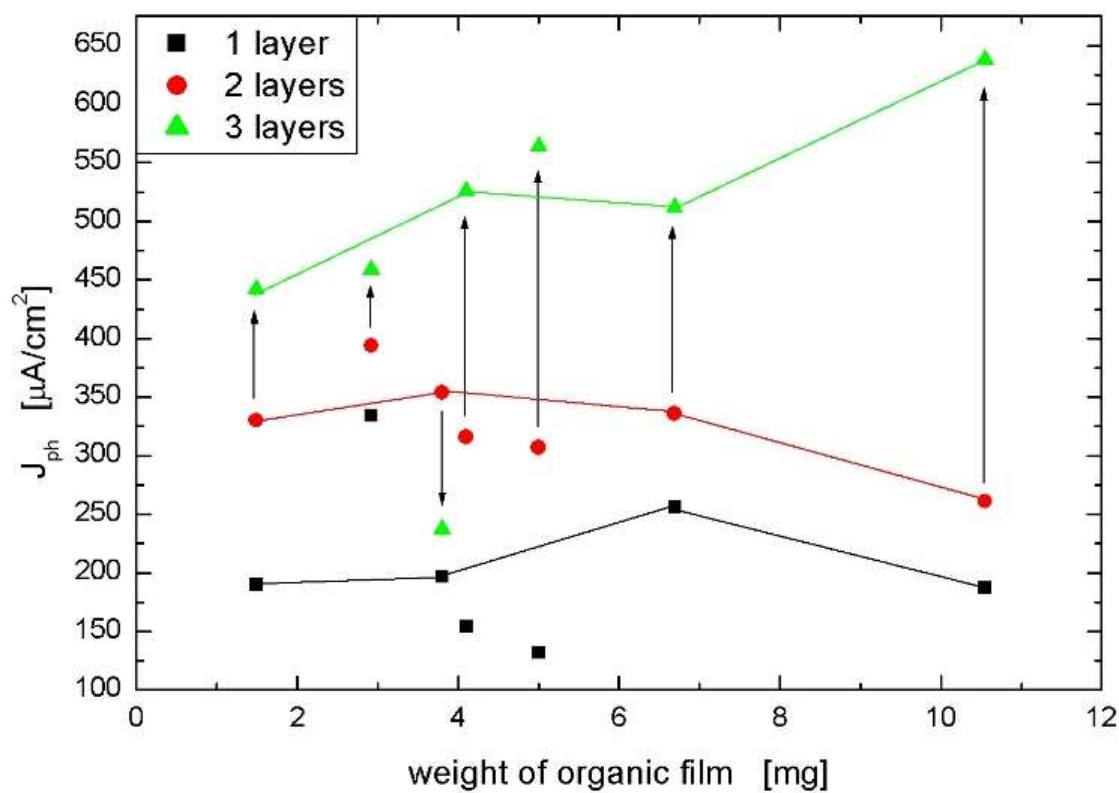

**Figure 14**2: Photocurrent density for one, two- and three-layer samples of precursor 4, the arrows indicate the change of photocurrent density for the third layer

The range of $R_{rms}$ is broadened to 600 nm to 1000 nm. For some samples the maximum peak height increases, causing a higher roughness value. But as could be seen for two-layer samples, the number of roughness peaks still increases. The thickness histogram of a three-layer sample shows higher frequencies of thicknesses especially higher than 1.5µm compared to the sample with two layers (see figure 15). The median is shifted to a range of 500nm to 700nm. The thickness histogram also reveals that the number of small thicknesses diminishes (see figure 15). The average is shifted to 800nm to 1000nm (see table 8). Therefore, thickness and roughness also increase for deposition of a third layer.

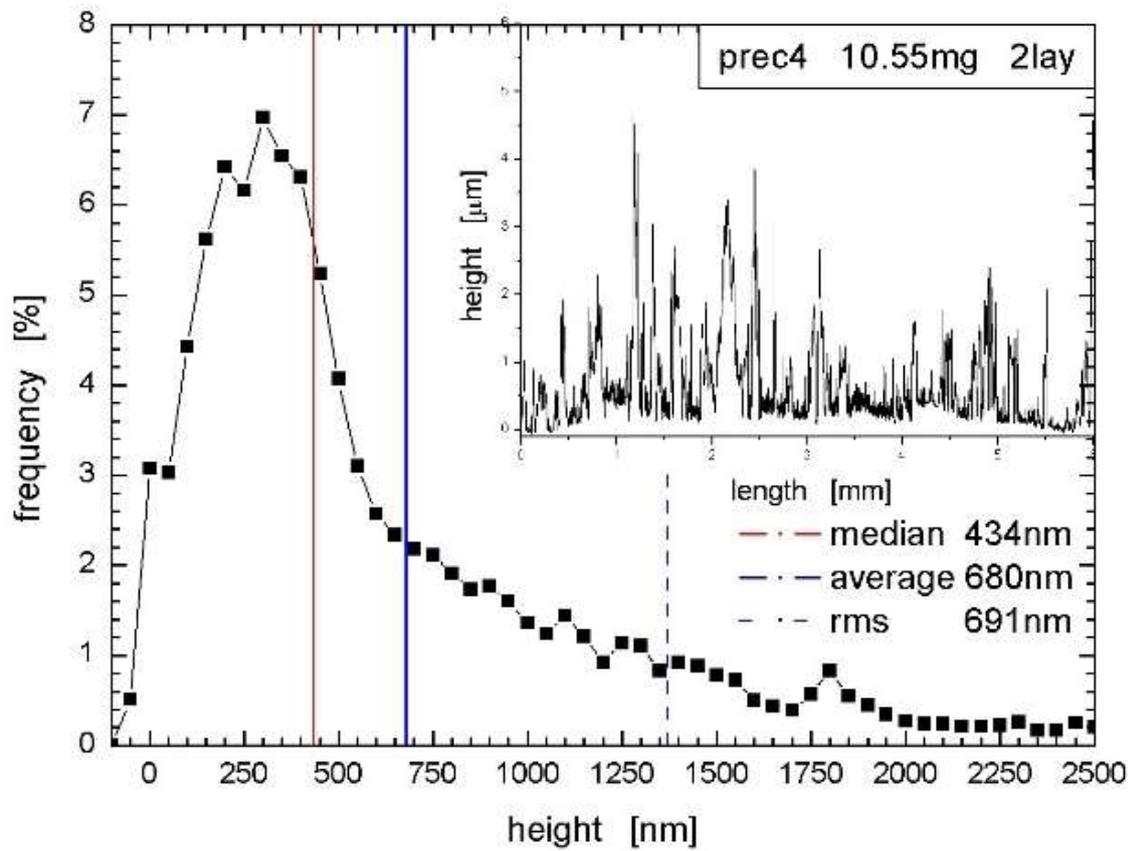

a) b)
27

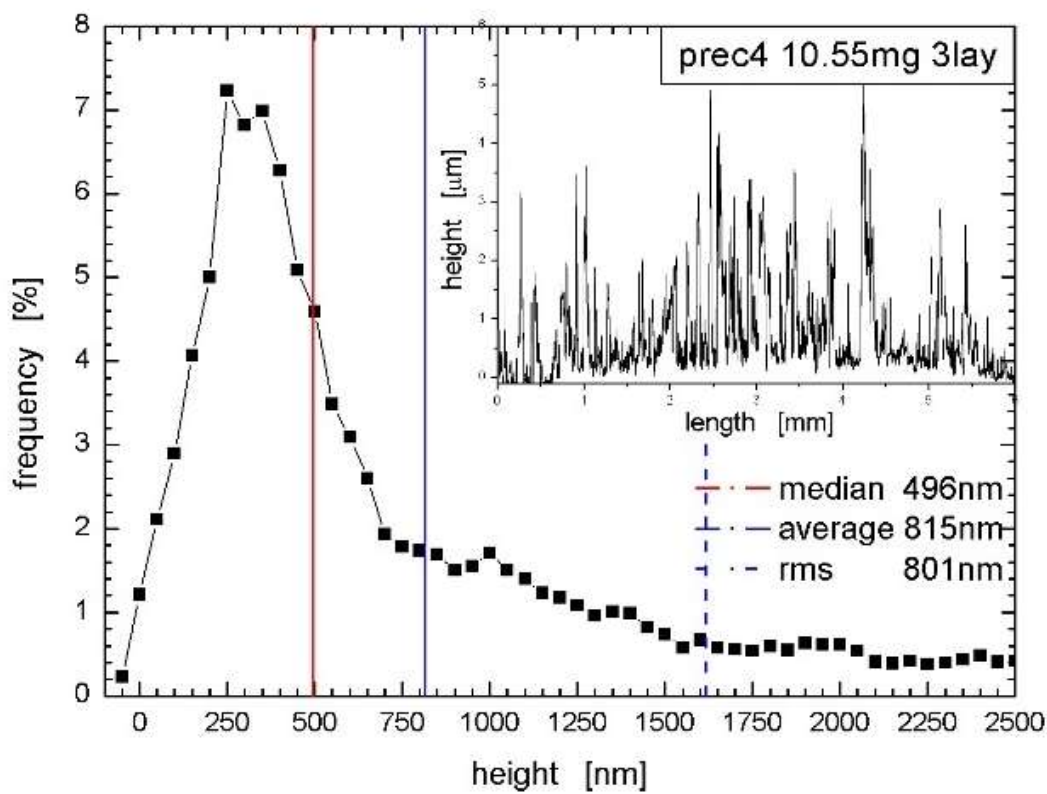

**Figure 3**: Thickness histograms and profiles for two-layer (a) and three layers (b) of sample preparation of precursor 4

From the single roughness and thickness values, no explanation can be found why the three-layer samples differ in PEC activity among each other. Also, there is no obvious reason for the decrease of photocurrent density of one sample. The sample of the highest photocurrent density shows the lowest median, 500nm.

In conclusion for a low thickness of the organic film (concentration studies), the hematite film is smooth and shows only a low PEC performance. With higher organic film thickness, the annealed film is rough, and the sample displays a high photocurrent density. With different precursor solutions different maximum PEC performance is achieved and variable roughness ranges could be the reason. Photocurrent density depends on the weight of organic film for one-layer samples, but this does not correlate with the measured thickness or roughness properties. A possible reason for this could be roughness on a scale that cannot be measured with the used settings for profilometry.

Roughness and film thickness as well as photocurrent density increase with deposition of further layers. Samples that show the highest values of photocurrent density display a roughness in the range 600nm≤$R_{rms}$≤800nm, but the $R_{rms}$ value provides limited information about the number of peaks that are in between this range. Bulk thickness up to 600nm to 700nm does not give a restriction to PEC performance. The different precursor solutions still provide different PEC performance so that for precursors with less performance, a higher number of layers must be deposited to achieve the same range of photocurrent density.

The roughness of samples seems to be one of the reasons for enhanced photocurrent density. It provides more room for better incident photon trapping, however the validity of this beyond the subject of the current study. Also, a scattering of light could be higher due to roughness and could, therefore, be used more effectively for water splitting.

### 3.2.1 PEC performance and thickness properties of spin-coated hematite thin films from an oleic acid derivative of an iron precursor

Spin coating was performed with solutions of precursors 3 and 4. With higher spin speed and less viscosity of the solution, a thinner organic film could be deposited. For deposition of the thin film, a slightly diluted solution and a highly concentrated solution (due to evaporation of solvent) with high viscosity of precursor 3 were used. Samples of the highly viscous solution show the same dependency of photocurrent density on the weight of organic film as dip coated samples (see figure 16a, red curve). Values of photocurrent density are shifted to a higher weight of organic film due to the accumulation of precursor at substrate edges. The thickness and roughness values coincide with the ranges of dip coated samples. For the slightly diluted precursor solution, both the values of photocurrent density and values of thickness and roughness are lower than for comparable dip coated samples (see figure 16a, green data points). Therefore, like for the dip coating process, a decrease of thickness, roughness, and photocurrent can also be seen for samples using a more diluted precursor solution. But in spin coating, a slight decrease in concentration already lowers the performance. Thus this method seems to be more affected by changes in the concentration of the precursor solution. Reason for the sensitivity of PEC performance to less concentration of used precursor solution could be the formation of the microstructure of organic film during evaporation of more solvent [15]. For deposition of precursor 4, the same precursor concentration was used as was for dip coating. PEC activity and thickness and roughness values do not differ from dip coated samples (see figure 16b).

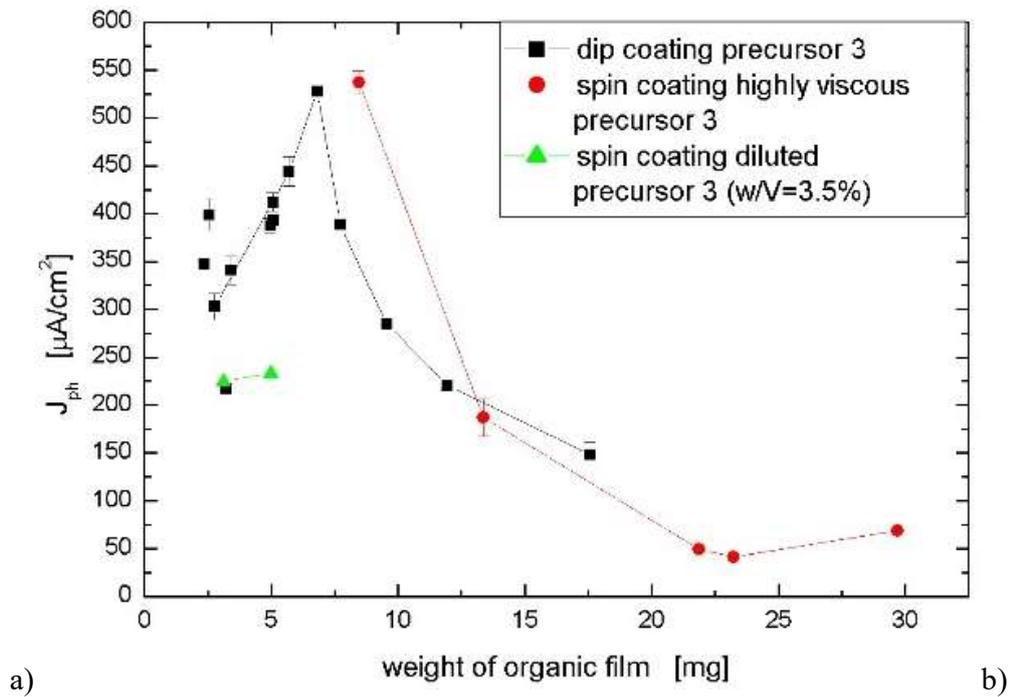

a)

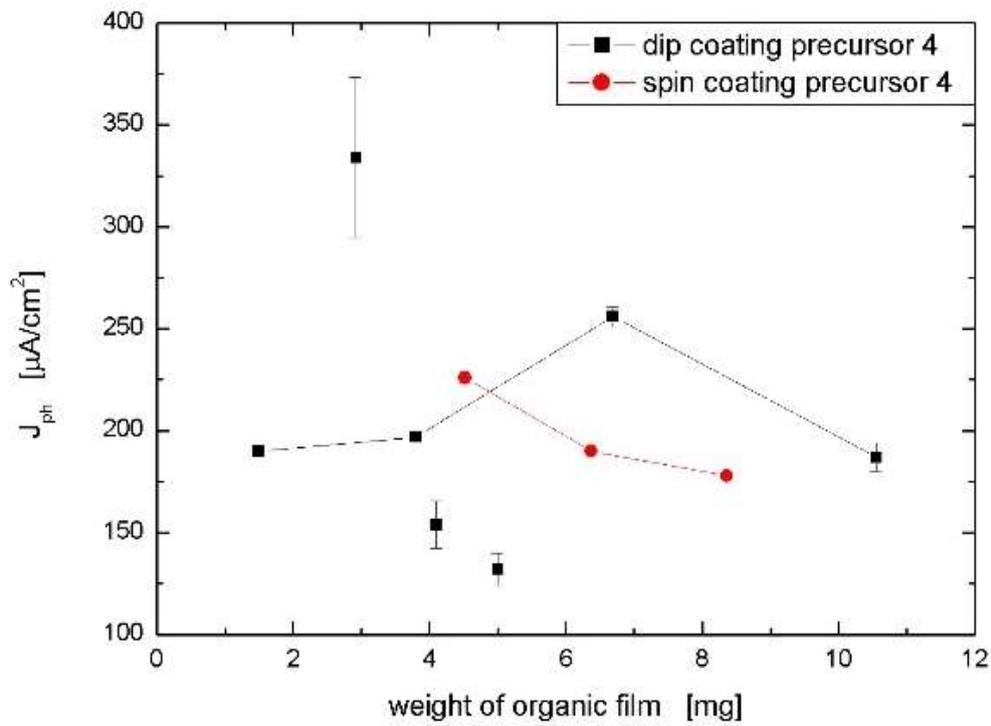

**Figure 16**: Comparison of achieved values of photocurrent density as a function of organic film weight for a dip and spin-coated samples of precursor 3 (a) and 4 (b)

## 3.3  PEC performance and thickness properties of dip-coated hematite thin films of stearic acid derivatives of iron precursor

Oleic acid was replaced by stearic acid for precursor preparation to investigate the influence of the different length of the fatty acid component. Samples were prepared using the dip coating process. During drying of dip coated samples mostly a bubble formation on the film could be observed. It indicates that reaction during preparation was not finished and the precursor still contains a trace amount of iron nitrate. Annealed samples with a weight of organic film up to 2mg show a smooth and compact hematite film with a thickness up to 500nm. With the higher weight of the organic film, much of the deposited material peels off after annealing so that the bare FTO substrate can be seen (see figure 17a). Thickness and roughness values of one-layer samples were not calculated as the high fraction of substrate in the profile would cause a significant error. Material that stayed shows high roughness peaks up to 10μm to 20μm (see figure 18 a).

PEC performance of samples is in the range of 10μA/cm$^2$ for the smooth sample to 50μA/cm$^2$ for rough samples (see figure 18b). The same dependency on the weight of organic film as for rough one-layer samples of an oleic acid derivative of the iron precursor can be found. A maximum in photocurrent density is located at organic film weights between 5 mg and 7mg (see figure 18 b). Samples of the longer heat-treated precursor solution show slightly bigger values of photocurrent density (see red data points figure 18 b). Also, for the current precursor, the PEC performance of samples increases with deposition of another layer (see figure 18b). For the two-layer sample, more material stays. The roughness and thickness values, as well as the photocurrent density, coincide with the ranges of rough one-layer samples prepared of iron oleate precursor 2.

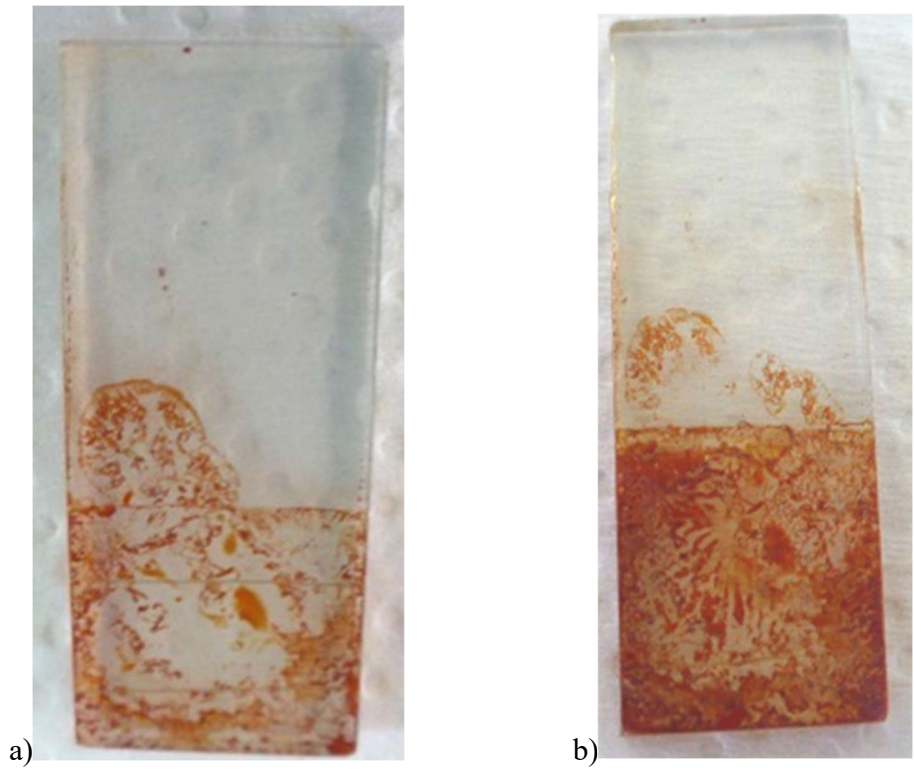

**Figure 17**: Peeled off the film of rough one (a) and two (b) layer dip coated sample of stearic acid precursor

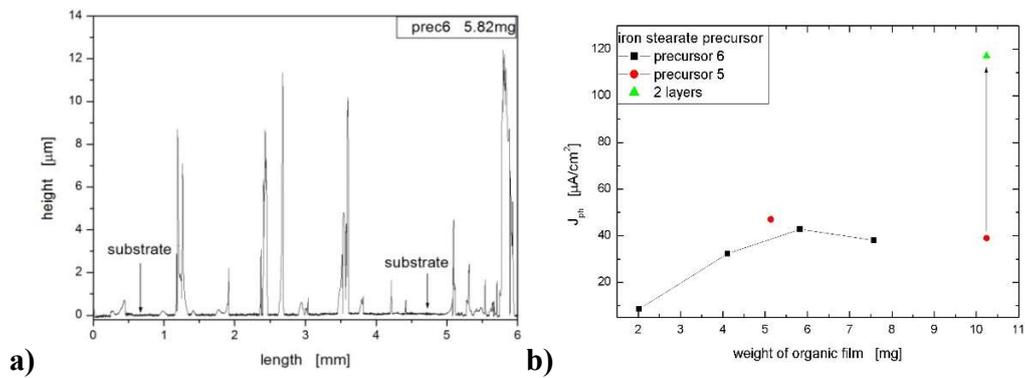

**Figure 18**: One-layer profile (a) and achieved photocurrent densities as a function of the weight of organic film (b) of dip coated samples of stearic acid precursor

## 3. 4. PEC performance and thickness properties of dip-coated hematite thin films of lauric acid derivatives of iron precursor

Lauric acid is used for precursor preparation to investigate the influence of a fatty acid with a shorter length of carbon chain. The reaction during the development of this precursor is less sluggish as for the stearic acid one but also here only foam and no viscous mass forms. Like for the stearic acid iron precursor, the achieved weight per volume concentration is low. A dip coating process prepared hematite film of lauric acid derivatives of the iron precursor. The annealed films show a combination of the smooth and rough surface but also areas where the material peels off completely and bare FTO substrate is left (see figure 19). It can also be seen in the profile in figure 19 a. From 0 mm to 2 mm of length, there is an area of high roughness, with a low median of 78 nm and arithmetic average and roughness values in the range of oleic acid derivatives of iron precursor 2. Maximum height peaks are around 8μm. Also, there is a smooth part of the film with an average thickness of 600 nm and roughness of 25 nm between 4mm and 6mm of length. Photocurrent densities of samples also display a dependency on the weight of organic film (see figure 20b). For a sample with a higher fraction of smooth surface the photocurrent density is low but increases to the range of 90μA/cm$^2$ to 110μA/cm$^2$ for rough samples with one layer.

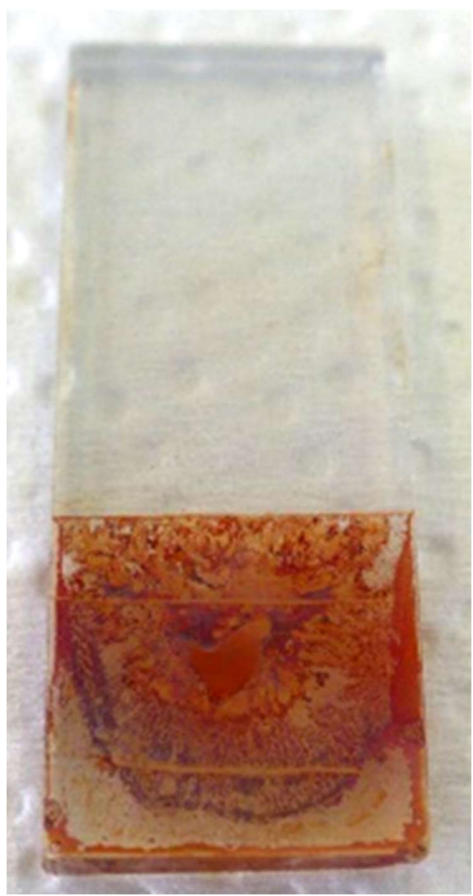

**Figure 19**: Dip coated sample of iron laurate precursor showing smooth (middle) and rough (around middle) hematite film and bare FTO substrate (bottom of the substrate)

a)

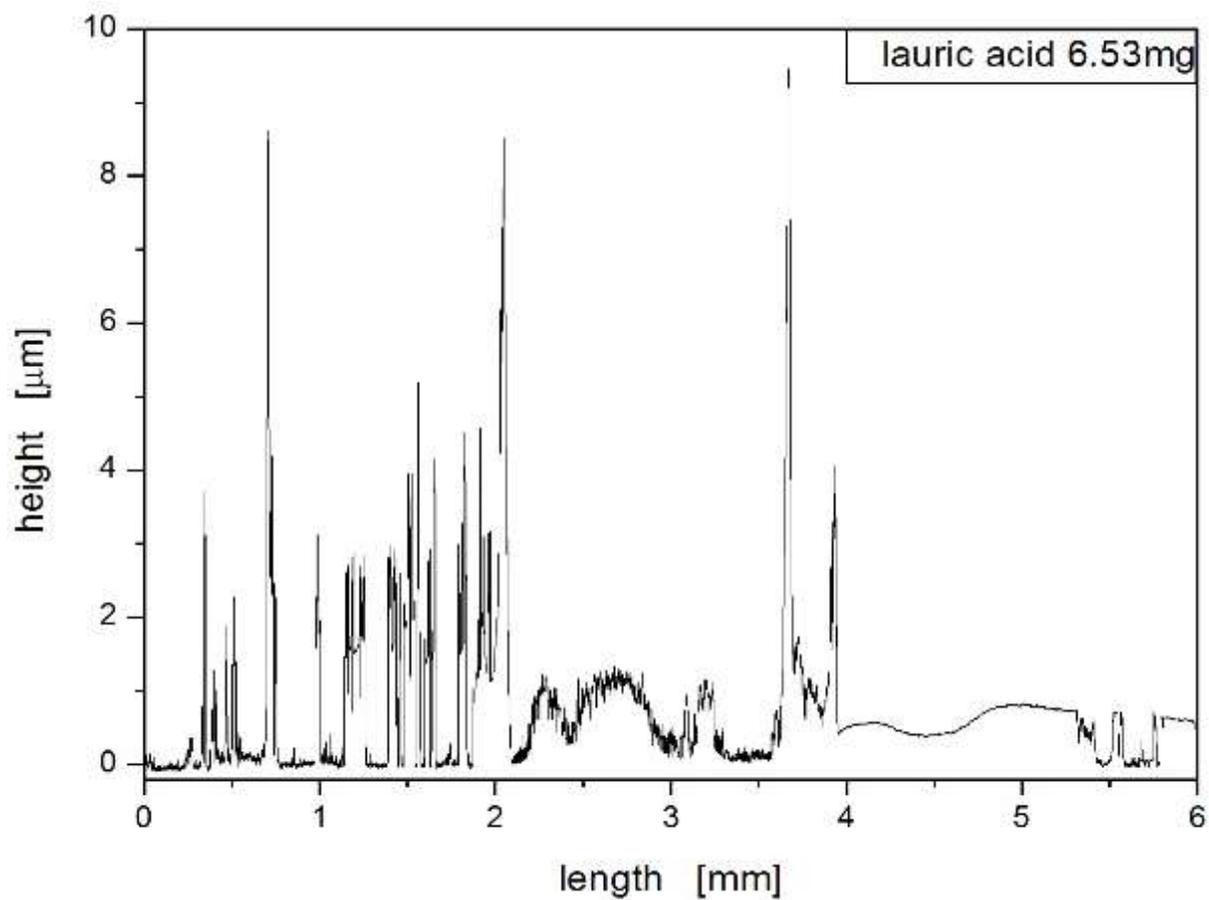

b)

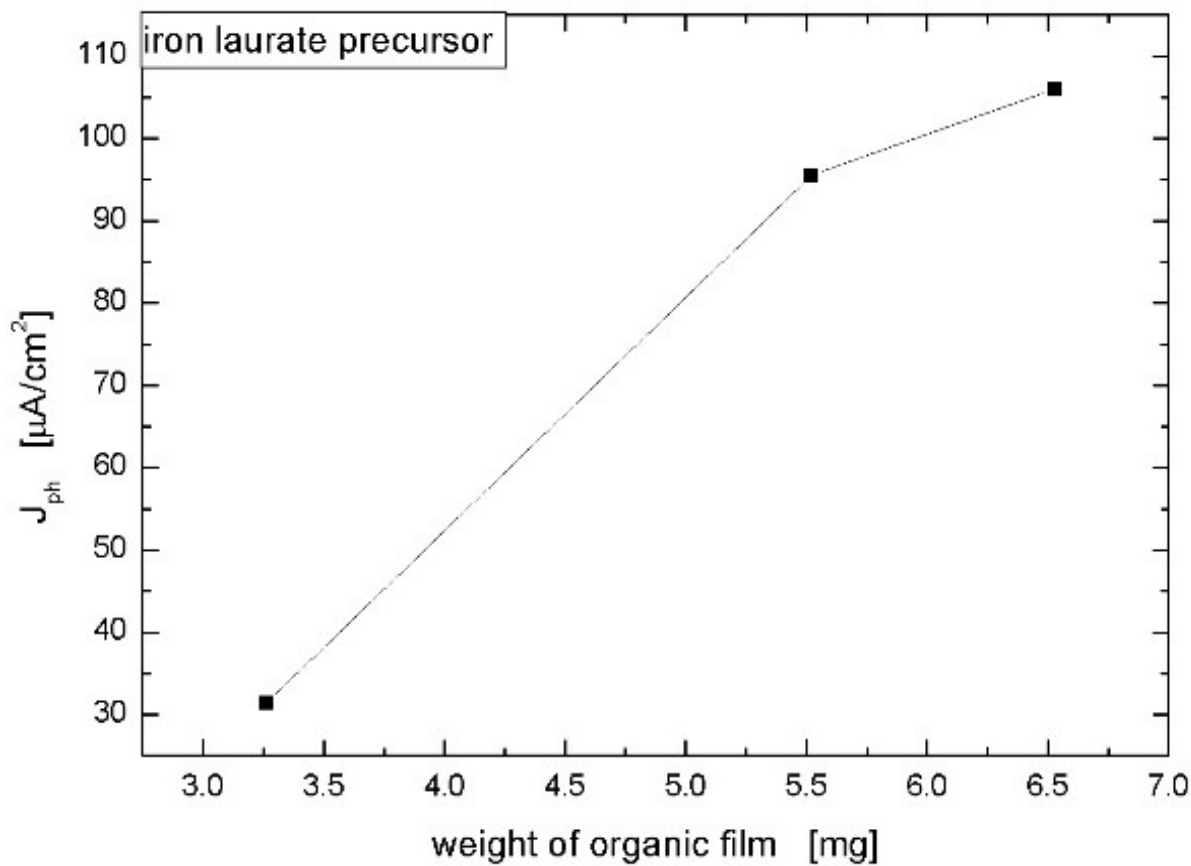

**Figure 20**: Thickness profile (a) and photocurrent density as a function of organic film weight (b) of samples of lauric acid derivatives of iron precursor

## 4. Conclusion

Different fatty acid derivatives of iron precursor were used to fabricate hematite photoelectrodes. It provides smooth hematite films for the low thickness of the organic film. Increasing the thickness of the organic film shows a rough surface on hematite after annealing. The rough surface is necessary for good PEC performance in case of these precursor. Different composition of one of the fatty acid derivative (iron-oleic acid) precursor solutions show a slight difference in decomposition behavior and lead to varying ranges of roughness and PEC performance. Photocurrent density values up to 0.6mA/cm$^2$ were achieved. Samples of good PEC performance show a roughness R$_{rms}$ in the range of 600nm to 800nm. For one-layer samples, photocurrent density depends on the thickness of organic film before annealing. This dependency could not be linked to the measured thickness and roughness values of annealed samples. Deposition of further layers on the annealed sample increases thickness, roughness, and PEC performance. Bulk thickness up to 600nm to 700nm sets no conductivity restriction on PEC performance. It is confirmed that the rough surface along with good micro porosity of the film, as found in the SEM investigation in the part two of the manuscript provides a better PEC performance.

**Acknowledgment**

Swiss Federal Office of Energy Project No. 100411 and the Swiss National Science Foundation R'Equip No. 206021-121306 are acknowledged for support of this research work.

**References**

1. United Nations Department of Economic and Social Affairs / Population Division: World Population Prospects: The 2008 Revision. In: *Population Newsletter*, Number 87, June 2009, pp. 1 – 4.
2. Energy Information Administration / International Energy Outlook 2009: International Energy Analysis Highlights to 2030 from:
    *http://www.eia.doe.gov/oiaf/ieo/pdf/highlights.pdf*
3. Bequerel, E.: Recherches sur les effets de la radiation chimique de la lumière solaire, au moyen des courants électriques. In: *Comptes Rendus de l'Académie des sciences* 9, 1839, pp. 145 – 149.
4. Miller, E. L.: Solar Hydrogen Production by Photoelectrochemical Water Splitting: The Promise and Challenge. In: Vayssieres, L. (Ed.): *On Solar Hydrogen & Nanotechnology*. Singapore: Wiley, 2009.
5. Bockris J: The origin of ideas on a Hydrogen Economy and its solution to the decay of the environment. *International Journal of Hydrogen Energy* 2002, 27: 731-740.
6. Bockris JOM: The hydrogen economy: It's history. *International Journal of Hydrogen Energy* 2013, 38: 2579-2588.
7. Gregory DP, Ng DYC, Long GM: The Hydrogen Economy. In *Electrochemistry of Cleaner Environments.* Edited by Bockris JO. Boston MA: Springer; 1972: 226-280.

8.  Lohaus C, Klein A, Jaegermann W: Limitation of Fermi level shifts by polaron defect states in hematite photoelectrodes. *Nat Commun* 2018, 9:4309.
9.  Fujishima, A.; Honda, K.: Electrochemical Photolysis of Water at a Semiconductor Electrode. In: *Nature* 238, 1972, pp. 37 – 38,
10. Lindgren, T.; Vayssieres, L.; Wang, H.; Lindquist, S.: Photo-oxidation of Water at Hematite Electrodes. In: Kokorin, A. I.; Bahnemann, D. W. (Eds.): *Chemical Physics of Nanostructured Semiconductor.* Boston: VSP, 2003.
11. Satsangi, V. R.; Dass, S., Shrivastav, R: Nanostructured α-Fe2O3 in PEC Generation of Hydrogen. In: Vayssieres, L. (Ed.): *On Solar Hydrogen & Nanotechnology.* Singapore: Wiley, 2009.
12. Sartoretti, C. J.; Ulmann, M.; Alexander, B. D. et al.: Photoelectrochemical oxidation of water at transparent ferric oxide film electrodes. In: *Chemical Physics Letters* 376, 2003, pp. 197 – 200.
13. Sartoretti, C. J.; Alexander, B. D.; Solarska, R. et al.: Photoelectrochemical oxidation of water at transparent ferric oxide film electrodes. In: *Journal of Physical Chemistry B*, 109, 2005, pp. 13685 – 13692.
14. Duret, A.; Grätzel, M.: Visible light induced water oxidation on mesoscopic α-Fe2O3 films made by ultrasonic spray pyrolysis. In: *Journal of Physical Chemistry B* 109, pp. 17184 – 17191.
15. Yarahmadi, S. S.; Tahir, A. A.; Vaidhyanathan, B.; Wijayantha, K. G. U.: Fabrication of nanostructured α-Fe2O3 electrodes using ferrocene for solar hydrogen generation. In: *Materials Letters* 63, 2009, pp. 523 – 526.
16. Souza, F. L.; Lopes, K. P.; Nascente, P. A. P.; Leite, E. R.: Nanostructured hematite thin films produced by spin-coating deposition solution: Application in water splitting. In: *Solar Energy Materials and Solar Cells* 93, 2009, pp. 362 – 368.
17. Spray, R. L.; Choi, K.-S.: Photoactivity of Transparent Nanocrystalline Fe2O3 Electrodes Prepared via Anodic Electrodeposition. In: *Chemistry of Materials* 21, 2009, pp. 3701 – 3709.
18. Brinker, C. J.; Hurd, A. J.; Schunk, P. R.; Frye, G. C.; Ashley, C. S.: Review of sol-gel thin film formation. In: *Journal of Non-Crystalline Solids* 147/148, 1992, pp. 424 – 436.
19. Scriven, L. E.: Physics and applications of dip coating and spin coating. In: *Materials Research Society Symposium Proceedings*, Vol. 121, 1988.

20. Larson, R. G.; Regh, T. J.: Spin coating. In: Kistler, S. F.; Schweizer, P. M. (Eds.): *Liquid film coating. Scientific principles and their technological implications*. London: Chapman & Hall, 1997.

21. Meyerhofer, D.: Characteristics of resist films produced by spinning. In: *Journal of Applied Physics* 49, 1978, pp. 3993 – 3997.

22. Debajeet K. Bora, Artur Braun, Selma Erat, Ahmad K. Ariffin, Romy Loehnert, Kevin Sivula, org T€opfer, Michael Gratzel, Ricardo Manzke, Thomas Graule, and Edwin C. Constable, J. Phys. Chem. C 2011, 115, 5619−5625.

23. Deb, P.; Basumallick, A.: Preparation of γ- Fe2O3 nanoparticles from a nonaqueous precursor. In: *Journal of Materials Research* 16, 2001, pp. 3471 – 3475.

24. Bronstein, L. M.; Huang, X.; Retrum, J.; Schmucker, A.; Pink, M.; Stein, B. D.; Dragnea, B.: Influence of Iron Oleate Complex Structure on Iron Oxide Nanoparticle Formation. In: *Chemical Materials* 19, 2007, pp. 3624 – 3632.

25. standard ISO 9845-1, 1992.

26. Bodschwinna, H.; Hillmann, W.: *Oberflächenmesstechnik mit Tastschnittgeräten in der industriellen Praxis*. Berlin, Köln: Beuth, 1992.

27. Gadalla, A. M.; Yu, H.: Thermal decomposition of Fe(III) nitrate and its aerosol. In: *Journal of Materials Research* 5, No. 6, 1990, pp. 1233 – 1236.